\documentclass[prd,showpacs,showkeys]{revtex4}
\usepackage{amsmath,amsthm,amsfonts,amssymb}
\usepackage[mathcal]{eucal}
\usepackage{mathrsfs}
\usepackage{graphicx}
\usepackage{dcolumn}
\usepackage{latexsym}
\usepackage{epsfig}
\usepackage{bm}
\usepackage[all]{xy}
\newcommand{\ie}{\begin{equation}}
\newcommand{\fe}{\end{equation}}

\def\text#1{\mbox{#1}}

\begin{document}

\title{{Scalar and tensor gauge field localization on deformed thick branes}}

\author{W. T. Cruz,$^{a,b}$ M. O. Tahim,$^{a,c}$  and C. A. S. Almeida$^a$}
\affiliation{$^a$Departamento de F\'{i}sica - Universidade Federal do Cear\'{a} \\ C.P. 6030, 60455-760
Fortaleza-Cear\'{a}-Brazil}
\affiliation{$^b$Centro Federal de Educa\c{c}\~{a}o Tecnol\'{o}gica do Cear\'{a} (CEFETCE), Unidade Descentralizada de Juazeiro do Norte, 63040-000 Juazeiro do Norte-Cear\'{a}-Brazil}
\affiliation{$^c$Departamento de Ci\^{e}ncias da Natureza, Faculdade de Ci\^{e}ncias, Educa\c{c}\~{a}o e Letras
do Sert\~{a}o Central (FECLESC), Universidade Estadual do Cear\'{a}, 63900-000
Quixad\'{a}-Cear\'{a}-Brazil}

\begin{abstract} We make an analysis about several aspects of localization
of a scalar field and a Kalb-Ramond gauge field in a specific four
dimensional AdS membrane embedded in a five dimensional space-time.
The membrane is generated from a deformation of the $\lambda \phi^4$
potential and belongs to a new class of defect solutions. The study
of deformed defects is important because they contain internal
structures and these may have implications to the way the background
space-time is constructed and the way its curvature behaves. The main
objective is to observe the contributions of the deformation
procedure to the well known field localization methods.
\end{abstract}




\maketitle

\section{Introduction}

In a scenario of extra dimensions the observable universe is represented by a four-dimensional membrane embedded in a higher dimensional space-time. The standard model of particles is confined in the membrane while gravitation is free to propagate into the extra dimension. These ideas have appeared as alternatives to solve the gauge hierarchy problem \cite{RS}. Recently a lot of attention has been given to the study of topological defects in the context of warped space-times. The number of extra dimensions guide us in choosing the right type of defect in order to mimic our brane-world. The key idea for construction of the brane-world is to localize in a very natural way the several fields  of our universe (the bosonic ones and fermionic ones). In this way several works have considered five-dimensional universes \cite{gremm,de} where five dimensional gravity is coupled to scalar fields. In this scenario, with a specific choice for the scalar potential, it is obtained thin domain wall as solutions that may be interpreted as non-singular versions of the Randall-Sundrum scenario. Besides gravity, the study of localization of fields with several spins it is very important \cite{kehagias}. Also, this type of scenario contributes for discussions about cosmology. In models with 5-dimensional membranes, the mechanism controlling the expansion of the universe have been associated to the thickness of the membrane along the extra dimension \cite{observations}.

As known, the kind of structure of the considered membrane is very
important and will produce implications concerning the methods of
field localization. In the works \cite{deformed,aplications} a class
of topological defect solutions is constructed starting from a
specific deformation of the $\phi^4$ potential. These new solutions
may be used to mimic new brane-worlds containing internal structures
\cite{aplications}. Such internal structures have implications in the
density of matter-energy along the extra dimensions \cite{brane} and
this produces a space-time background whose curvature has a
splitting, as we will show, if compared to the usual models. Some
characteristics of such model were considered in phase transitions
in warped geometries \cite{fase}.

Motivated by the references above, our main subject here it is to answer the
following question: Are these structures able to localize fields of
variable spins? In order to find the answer, in this work we propose
the analysis of a real scalar field and the gauge tensor field, or
Kalb-Ramond field. The paper is organized as follows: in the second
section we describe how the deformed membrane is constructed and how
the space-time background is obtained; in the third section we study
the localization of massless and massive modes of a scalar field in
the background obtained; the fourth section treats the behavior of
the Kalb-Ramond gauge field by using the same basic steps; the
fifth section is important because we introduce the dilaton field in
order to force the localization of the Kalb-Ramond field. Such analysis is made in the sixth section. At the final we present our
conclusions and perspectives.

\section{Two-kink solutions modeling the brane}

There is great interest in studying scalar fields coupled to
gravity. If we consider a $D=5$ universe, we should embed a kink
solution in this space-time in order to build our membrane. These
kind of solutions are obtained through the $\lambda\phi^4$ or
sine-Gordon potentials. In our case, following the reference
\cite{deformed}, we will obtain a new class of defects starting from
a deformation of the $\lambda\phi^4$ potential. In this way we can
analyze localization of fields of several ranks in a more complete
fashion because the deformed membranes suggests the existence of
internal structures. As we will see, this choice avoids space-time
singularities also, which is only possible by choosing smooth membrane
solutions.

Our model is built with an AdS $D=5$ space-time whose metric is
given by
\begin{equation}
ds^{2}=e^{2A(y)}\eta_{\mu\nu}dx^{\mu}dx^{\nu}+dy^{2}.
\end{equation}
The warp factor is composed by the function $A(y)$, where $y$ is the
extra dimension. The tensor $\eta_{\mu\nu}$ stands for the Minkowski
space-time metric and the indexes $\mu$ and $\nu$ go from $0$ to
$3$.

In order to construct the membrane solution we start with an action
describing the coupling between a real scalar field and gravitation:
\begin{equation}
S=\int d^{5}x
\sqrt{-G}[2M^{3}R-\frac{1}{2}(\partial\phi)^{2}-V(\phi)].
\end{equation}
In the last action, the field $\phi$ represents the stuff from which
the membrane is made, $M$ is the Planck constant in $D=5$ and $R$ is
the  scalar curvature. The equations of motion coming from that
action are:
\begin{equation}\label{mov1}
\frac{1}{2}(\phi^{\prime})^{2}-V(\phi)=24M^{3}(A^{\prime})^{2},
\end{equation}
and
\begin{equation}\label{mov2}
\frac{1}{2}(\phi^{\prime})^{2}+V(\phi)=-12M^{3}A^{\prime\prime}-24M^{3}(A^{%
\prime})^{2}.
\end{equation}
Note that the prime means derivative with respect to the extra
dimension. Basically, we look for solutions in which $\phi$ tends to
different values when $y\rightarrow\pm\infty$. In a flat space-time
we find kink-like solutions for the above equations by choosing a
double-well potential. Analogously, if we look for bounce-like
solutions in curved space-time, we should regard potentials
containing various minima. In the presence of gravity, we can find
first order equations by the superpotential method if we take the
superpotential $W(\phi)$ in such a way that $\frac{\partial
W}{\partial \phi}=\phi'$. Our potential must be defined by
\begin{equation}
V_p(\phi)=\frac{1}{2}\left(\frac{dW}{d\phi}\right)^2-\frac{8M^3}{3}W^2,
\end{equation}
from where we can conclude that $W=-3A'(y)$. This formalism was
initially introduced to study non-supersymmetric domain walls in
various dimensions \cite{de,sken}.

Following the references \cite{deformed,aplications,adauto,defects-inside}
the superpotential is given by,
\begin{equation}\label{sup}
W_p(\phi)=\frac{p}{2p-1}\phi^{\frac{2p-1}{p}}-\frac{p}{2p+1}\phi^{\frac{2p+1}{p}},
\end{equation}
where $p$ is an odd integer. The choice for $W_p$ can be obtained by
deforming the usual $\phi^4$ model and it is introduced in the study of
deformed membranes \cite{adauto}. This choice will permit us to get
new and well behaved models for $p=1,3,5,...$ (for $p=1$ we get the
usual $\phi^4$ model). For $p=3,5,7,...,$ the potential $V_p$ has
one minimum at $\phi=0$ and two at $\pm 1$. A new class of solutions
called two-kink solutions initially presented in Ref.\cite{aplications} can be
obtained from the choice of the superpotential $W_p$. For this we
solve $\frac{\partial W}{\partial \phi}=\phi'$ to find
\begin{equation}\label{twokink}
\phi_p(y)=tanh^p(\frac{y}{p}).
\end{equation}
Starting from the first order equation $W_p=-3A_p'(y)$, we can find
the solution for the function $A_p(y)$ \cite{adauto},
\begin{eqnarray}\label{a}
A_p(y)=-\frac{1}{6}\frac{p}{2p+1}\tanh^{2p}\left(\frac{y}{p}\right)-
\frac{1}{3}\left(\frac{p^2}{2p-1}-\frac{p^2}{2p+1}\right)\\\nonumber
\biggl{\{}\ln\biggl[\cosh\left(\frac{y}{p}\right)\biggr]-
\sum_{n=1}^{p-1}\frac1{2n}\tanh^{2n}\left(\frac{y}{p}\right)\biggr{\}}
\end{eqnarray}
The function $A(y)$ determines the behavior of the warp factor. The
characteristics of localization for several fields and the
construction of effective actions in $D=4$ will depend on part of
the contribution of the warp factor. Note that the exponential
factor constructed with this function is localized around the
membrane and for large $y$ it approximates the Randall-Sundrum
solution \cite{RS}. The solution found here reproduces the
Randall-Sundrum model in an specific limit. The space-time now has
no singularity because we get a smooth warp factor (because of this,
the model is more realistic) \cite{kehagias}. In fact this can be
seen by calculating the curvature invariants for this geometry. For
example, we obtain


\begin{figure}{\centerline{
\epsfig{figure=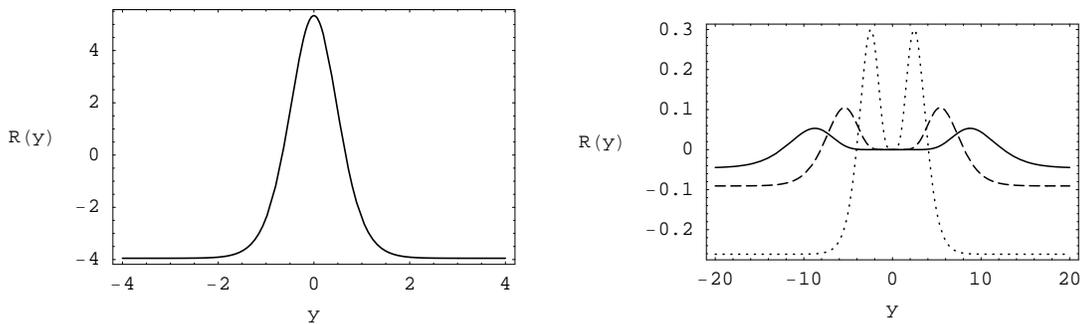,width=15.2cm,height=4.6cm}}
\caption{Plots of the solution of the curvature invariant
$R(y)$$p=1$ (left) and for $p=3$ (dashed line), $p=5$ (doted line)
and $p=7$ (solid line) (right) .}\label{curvp}}
\end{figure}

\begin{equation}
R=-[8A_p''+20(A_p')^{2}],
\end{equation}
Note that the Ricci scalar is finite, which we can observe through
the Fig.(\ref{curvp}). We can see also an important characteristic
of the deformed structure in comparison with the usual thick
membrane models generated by simple kinks. Observing again the
Fig.(\ref{curvp}), for $p=1$, we have the usual curvature scalar
for the non-deformed model, the usual one. In this case, the
curvature has maximum at $y=0$ and goes to negative values when
$y\rightarrow\infty$. However, regarding the deformed model, taking
$p=3,5$ in $R_p(y)$, we obtain a splitting with the appearance of a
region of zero curvature between two maxima.

\section{Scalar field}

In this section we first analyze the localization of zero-modes for
a spin zero (or scalar) field in the background described in the
first section. After that we study the spectrum of massive modes for
the scalar field in the context of deformed thick brane. It is important
to note that our thick brane solution $\phi_p$ was built from a
kink-like solution. We now analyze the localization of the scalar
field $\Phi$ in the membrane described above. We find new aspects
for localization on this new background.

\subsection{Localized zero-mode}

Starting from the membrane discussed in the last section  we will
analyze the localization of a real scalar field. A similar work can
be found in Ref.\cite{borut}, where a zero mode is studied in a more
basic scenario. The deformation give us more details, especially towards the study of field localization. The potential and the wave
function are modified when we modify the way the background
interacts with the fields of this model. These new solutions show us
new behaviors related to resonances and the mass spectrum.

Initially we take the action for the scalar field $\Phi$ coupled to
gravity,
\begin{equation}\label{acao}
\frac{1}{2}\int d^4xdy \sqrt{-G}G^{MN}\partial_M\Phi\partial_N\Phi,
\end{equation}
where the indexes $M,N$ go from $0$ to $4$. We continue by finding
the equations of motion,
\begin{equation}\label{eqmov}
\eta^{\mu\nu}\partial_\mu\partial_\nu\Phi+e^{-2A_p(y)}\partial_y[e^{4A_p(y)}\partial_y\Phi]=0,
\end{equation}
where we have separated the extra dimension from the
others. In the next step we use the following separation of
variables:
\begin{equation}\label{eq}
\Phi(x,y)=\chi(x)\psi(y).
\end{equation}
Writing $\Phi$  as above we arrive at the following equation for the
$y$ dependence:
\begin{equation}
4A_p'\frac{d\psi}{dy}+\frac{d^2\psi}{dy^2}=-m^2
e^{-2A_p}\psi.\label{ydep}
\end{equation}
This equation is very similar to the equation for gravity
localization \cite{RS}. For $m^2=0$ we see that $\psi(y)=c$ is a
solution, where $c$ is constant. The condition to obtain a localized
scalar zero mode from equation (\ref{ydep}) in this case is just to
find a behavior which can be interpreted as something trapped to the
membrane. In some sense, the solution $\psi(y)=c$ can not be
localized because it is not suppressed in regions far from the
membrane. However, the correct way to see localization is by
studying the effective action in $D=4$. Going back to the action
(\ref{acao}) we make again the procedure of separation of variables
and, from the relation (\ref{eq}) and from the result $\psi(y)=c$, we
get
\begin{equation}
\frac{1}{2}\int_{-\infty}^{+\infty} dy \psi^2 e^{2A_p}\int d^4x
\eta^{\mu\nu}\partial_\mu\chi\partial_\nu \chi.\label{ca1}
\end{equation}
The part dependent on the extra dimension in the action above will
be determined by the warp factor behavior resulting from this new
kind of membrane.

\begin{figure}{\centerline{
\epsfig{figure=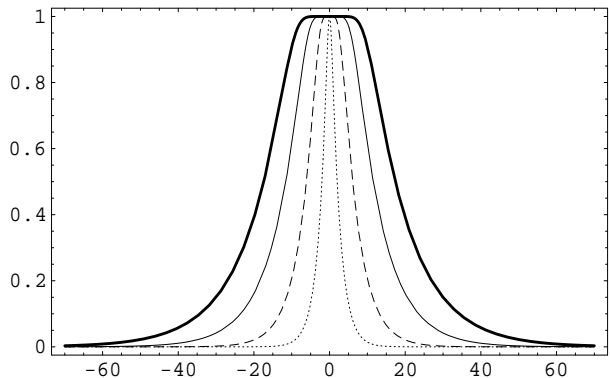,width=8cm,height=5cm}}
\caption{Plots of $\psi^2 e^{2A_p}$ for p=1 (points), p=3
(dashed), p=5 (line) and p=7 (thick).}\label{psip}}
\end{figure}

We can observe in Fig.(\ref{psip}) the behavior of the functions
important to localization of the effective action (\ref{ca1}). We make a plot
for several cases of $A_p$, $p=1,3,5,7$. From a first view, it seems
that there are good conditions to obtain localization. However, is
this result valid for all values of $p$? From Fig.(\ref{psip}), the
integral increases if we increase $p$. If we demonstrate the
convergence of the effective action for every $p$, we demonstrate
the existence of a localized zero mode for any function $A_p$.

\begin{figure}{\centerline{
\epsfig{figure=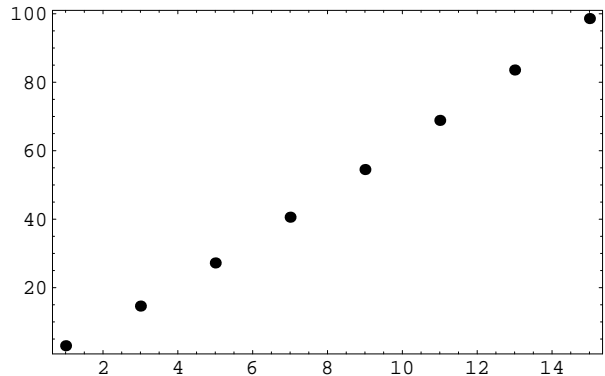,width=8cm,height=5cm}} \caption{Plots
of $\int_{-\infty}^{\infty}\psi^2 e^{2A_p}dy$ for various
$p$.}\label{p} }
\end{figure}

In Fig.(\ref{p}) we compute the evolution of values for the part
of the effective action in the extra dimension in terms of $p$. The
conclusion is that, given a solution for $\psi(y)$, for finite
values of $p$, the warp factor obtained give us localization on the
membrane. When $p\rightarrow\infty$ the warp factor delocalize the
zero mode. A similar feature occurs in the study of splitting of
thick branes due to phase transitions \cite{fase}. In this case, the
splitting delocalize the gravitational zero mode.

\subsection{Massive spectrum}

\begin{figure}{\centerline{
\epsfig{figure=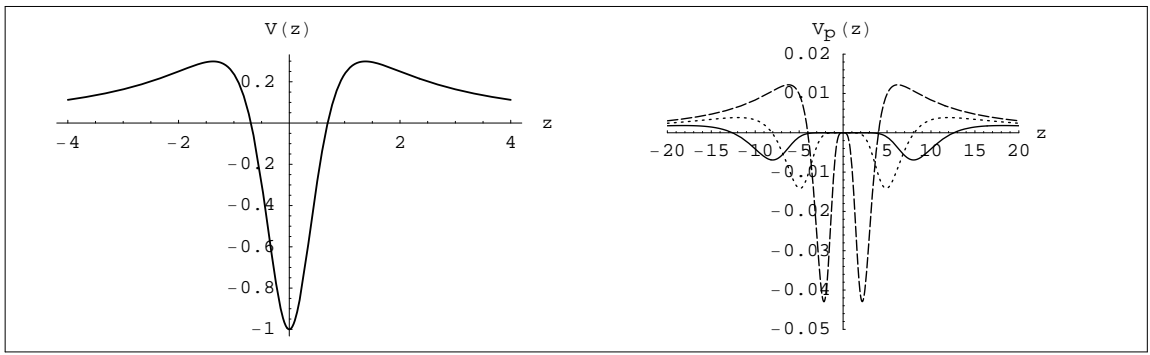,width=15.2cm,height=4.6cm}}
\caption{Plots of $\overline{V}_p(z)$ for $p=1$ (left) and
for $p=3$ (dashed line), $p=5$ (doted line) and $p=7$ (solid line)
(right) .}\label{volp}}
\end{figure}

In order to study the Kaluza-Klein states of the scalar field we
should rewrite its equation of motion (the part due to the extra
dimension) in a Schrodinger-like equation. For this, we get Eq.
(\ref{ydep}), i. e.,
\begin{equation}
\left\{\frac{d^2}{dy^2}+ 4A'\frac{d}{dy}\right\}\psi=-m^2
e^{-2A}\psi.
\end{equation}
and make the following transformation,
\begin{equation}\label{trans}
y\rightarrow z=f(y),\,\,\,\,\,\,  \psi=\Omega\overline{\psi}.
\end{equation}
The conditions to arrive in a Schrodinger-like equation tell us the
form of the function $\Omega$. In this way we discard terms of first
order in derivatives and the right hand side of this equation
contains the squared mass $m^2$. Then we obtain
\begin{equation}
\Omega=e^{-\frac{3}{2}A}, \,\,\,\,\,\, \frac{dz}{dy}=e^{-A}.
\end{equation}
Using the transformations (\ref{trans}) the Schrodinger equation can be written as,
\begin{equation}\label{schrop}
\left\{-\frac{d^2}{dz^2}+\overline{V}_p(z)\right\}\overline{\psi}=m^2\overline{\psi},
\end{equation}
where the potential $\overline{V}_p(z)$ assumes the form
\begin{equation}
\overline{V}_p(z)=e^{2A_p}\left[\frac{15}{4}(A_p')^2+\frac{3}{2}A_p''\right].
\end{equation}
We can write the potential in terms of derivatives of the $z$
variable, i. e.,
\begin{equation}
\overline{V}_p(z)=\left[\frac{9}{4}(\dot{A_p})^2+\frac{3}{2}\ddot{A_p}\right]
\end{equation}
We plot the potential in Fig.(\ref{volp}) for $p=3,5$ e $7$. The
figure strongly suggests the presence of resonances in the massive
spectra. Obviously, for $p=1$ we reobtain the structures of the
volcano potential. More interesting details occurs when $p=3,5,7...$
and we can clearly observe the presence of internal structures
\cite{adauto}. More specifically, for $p=1$ we have only
one minimum at $z=0$.  By changing the values of $p$, the potential
presents two minima and this is interpreted as the appearance of
internal structures. Following Campos \cite{fase} reasoning about the potential in the Schrodinger equation in the transverse and traceless sector of metric fluctuations, the appearance of these characteristics are due to a
phase transition.

The asymptotical behavior of the potential give us information about
the presence of gaps in the spectrum. As we can observe,
$\overline{V_p}(z)\rightarrow 0$ when $z\rightarrow \infty$. This
excludes the possibility of gaps. Is easy to note that the
Schrodinger equation (\ref{schrop}) may be written in a
supersymmetric quantum mechanical form:
\begin{equation}\label{qm2}
Q^{\dag} \, Q \, \overline{\psi}(z)=\left\lbrace
-\frac{d}{dz}-\frac{3}{2}\dot{A}_p\right\rbrace \left\lbrace
\frac{d}{dz}-\frac{3}{2}\dot{A}_p\right\rbrace \overline{\psi}(z)=
m^2\overline{\psi}(z).
\end{equation}
In this way we can avoid the existence of tachyonic modes, a
necessary condition to keep the stability of the gravitational
background discussed here.

As is pointed out in Refs. \cite{gremm} and \cite{csaba}, for highly
massive modes in relation to $\overline{V}(z)_{max}$, the potential
represents only a little perturbation. Nevertheless, it is possible
that modes of the function $\overline{U}(z)$ for which
$m^2\leq\overline{V}(z)_{max}$ can resonate with the potential. In
order to investigate this possibility it is important to study the
wave function $\overline{\psi}$, for several eigenvalues $m^2$, from
the equation (\ref{schrop}). We can solve numerically this equation
in order to better understand the presence of resonances.

\begin{figure}{\centerline{
\epsfig{figure=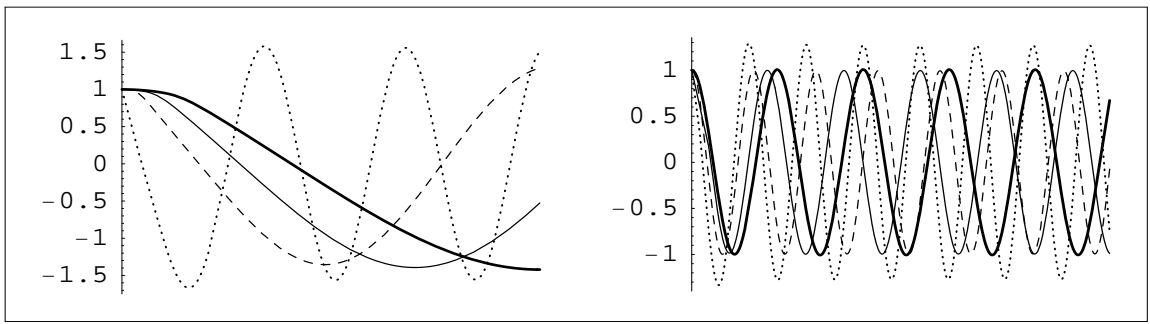,width=15.2cm,height=4.2cm}}
\caption{
 Plots of $\overline{V}_p(z)$ for $p=1$ (doted line), $p=3$ (dashed
line), $p=5$ (solid line) and $p=7$ (thick line), where we put $m^2
\leq {\overline{V}_p}(z)_{max}$ (left) and  $m^2 >
{\overline{V}_p}(z)_{max}$ (right)}\label{wavep}}
\end{figure}

Given the potential we should observe the appearance of resonances
when $m^2 \leq {\overline{V}_p}(z)_{max}$. The principal signal is
that inside the membrane the resonances must have huge amplitude
oscillations. With the function $A_p$  and the potential we plot in
Fig.(\ref{wavep}) several solutions of the equation
(\ref{schrop}) for some values of $m^2$ e $p$. As we can see, the
wave function oscillates quickly for a moderate value of $m^2$ and
it reduces its period for small values of $m^2$. Studying the
solutions for the wave function we have not found any evidence of
resonances. As mentioned in reference \cite{adauto}, the
behavior of the wave function suggests freedom on the bulk without
trapping on the membrane.

\begin{figure}{\centerline{ \epsfig{figure=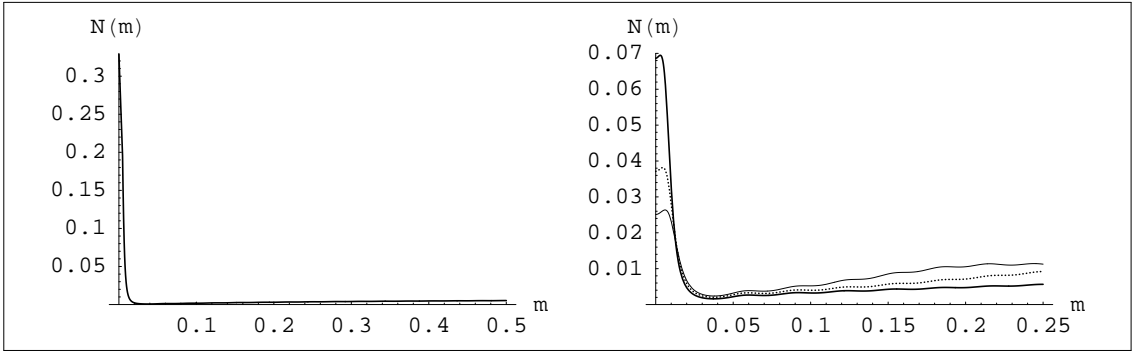,width=15cm}} \caption{
 Plots of $N_p(m)$ for $p=1$ (left), $p=3$ (thick line), $p=5$ (points) and $p=7$ (thin line) (right).}\label{re}}
\end{figure}

In order to enlarge our understanding about the coupling between the zero modes and matter inside the membrane we go back to the equation (\ref{qm2}).
The quantity $|\chi\overline{\psi_p}(z)|^2$ may be interpreted as the probability of finding the mode in the position $z$, where $\chi$ is a
normalization constant \cite{ca,csaba2}. In this way, by computing $|\chi\overline{\psi_p}(0)|^2$ we will know the probability to find the mode in
the membrane. For this, we solve numerically Eq.(\ref{schrop}) again, from where we just extract the values $\overline{\psi}_p(0)$ as functions of the
mass. We have limited ourselves to find solutions of Eq.(\ref{schrop}) satisfying $0<m<1$ because if $m^2>>V_{max}$  the potential will only represents
a small perturbation \cite{gremm}. In order to normalize the plane wave function we have limited the solutions for each mode to the region
$-100<z<100$ and have extracted a normalization constant $\chi$ for each mass value. With these results at hand we construct the function
$N_p(m)=|\chi\overline{\psi}_p(m)^2|$ which give us the probability of finding the modes at $z=0$ as function of $m$ for each solution $A_p(z)$ and
$\overline{\psi}_p$. In the Fig.(\ref{re}) we plot the function $N_p(m)$ and observe, for $p=1$, a peak of resonance for $m=0$. This structure is
similar to that obtained in Ref.\cite{csaba2} in the study of graviton massive modes. The resonance for $m=0$  is consistent with the results presented
in section 3.1 where we have obtained, via effective action, a localized zero mode for $p=1$. Considering the Schrodinger equation (\ref{schrop}),
the coupling of the modes with the matter on the membrane is related to the amplitude of the plane wave function normalized at $z=0$ \cite{gremm}. In
our case, we attribute this relation to the value of the function $N_p(m)$. The resonance for $p=1$ show us that the massive modes do couple weakly
with the membrane in comparison with the zero mode. However, observing the solutions of $N_p(m)$ for $p=3,5,7$ again in Fig.(\ref{re}) we verify
that the resonant structure for $m=0$ tends to disappear when we increase the values for $p$. This is related to the fact that we can only guarantee the existence of a zero mode for finite values of $p$.

\section{The Kalb-Ramond field}

In this section we analyze the behavior of the Kalb-Ramond field in the presence of membranes with internal structures. In this case we study
mechanisms of localization and normalization for its zero modes and for their Kaluza-Klein modes.

\subsection{Zero-mode}

Firstly we introduce in the action of the deformed membrane the
Kalb-Ramond field in the following way
\begin{equation}
\int d^{5}x \sqrt{-G}[2M^{3}R-\frac{1}{2}(\partial\phi_p)^{2}-V_p(%
\phi)-H_{MNL}H^{MNL}],
\end{equation}
where $H_{MNL}=\partial_{[M}B_{NL]}$ is the field strength  for the
Kalb-Ramond field. We will make the gauge choice $B_{\alpha 5}=0$ in
such that its non null components are only those living in the
membrane. We have found the equations of motion for $B_{MN}$ and
made explicit the part dependent on the extra dimension:
\begin{equation}
e^{-2A_p}\partial_{\mu}H^{\mu\gamma\theta}-\partial_{y}H^{y\gamma\theta}=0.
\end{equation}
We make now a separation of variables in order to work the part of
the extra dimension,
\begin{equation}
B^{\mu\nu}(x^{\alpha},y)=b^{\mu\nu}(x^\alpha)U(y)=b^{\mu\nu}(0)e^{ip_\alpha
x^\alpha}U(y),
\end{equation}
where $p^2=-m^2$. We write $H^{MNL}$ as $h^{\mu\nu\lambda}U(y)$. The
equation of motion becomes:
\begin{equation}
\partial_{\mu} h^{\mu\nu\lambda}U(y)-e^{2A_p}\frac{d^2U(y)}{dy^2}b^{\nu\lambda}e^{ip_\alpha x^\alpha}=0.
\end{equation}
The function U(y) carry all information about the extra dimension
and obeys the following equation:
\begin{equation}\label{motion}
\frac{d^2U(y)}{dy^2}=-m^2e^{-2A_p(y)}U(y)
\end{equation}
When $m^2=0$ we have the solutions $U(y)=cy + d$ and $U(y)=c$ with
$c$ and $d$ constants. With these at hand we start to make
computations in order to find localized zero modes of the
Kalb-Ramond field in the deformed membrane. We take the effective
action for the tensor field where we decomposed the part dependent on
the extra dimension,
\begin{equation}
S\thicksim\int d^{5}x \sqrt{-G}(H_{MNL}H^{MNL})= \int dy U(y)^2
e^{-2A_p(y)} \int d^4 x(h_{\mu\nu\lambda}h^{\mu\nu\lambda}).
\end{equation}\label{effec_action}
Given the solutions for $A_p$ and for $U(y)$ obtained above, we
clearly observe that due to the minus sign in the warp factor, the
function $U(y)^2 e^{-2A_p(y)}$ goes to infinity for the two
solutions of $U(y)$. In this way, the effective action for the zero
mode of the Kalb-Ramond field is not finite after integrating the
extra dimension.

\subsection{Massive modes}

\begin{figure}{\centerline{ \epsfig{figure=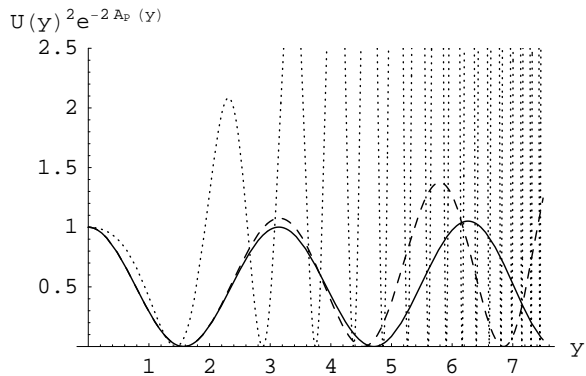,width=8cm,height=5cm}}
\caption{
 Plots of $U(y)^2 e^{-2A_p(y)}$ for $p=1$ (doted line), $p=3$ (dashed
line) and $p=5$ (solid line), where we put $m=2$.}\label{wavediv}}
\end{figure}

In order to make more complete our analysis of the behavior of the
tensor gauge field in this background we can take solutions for
$m\neq0$ in the equation (\ref{motion}). We must make the following
transformation,
\begin{equation}
y\rightarrow z=f(y),\,\,\,\,\,\,f'(y)=e^{-A_p}\,\,\,\,\,\,
U(y)=e^{\frac{1}{2}A_p}\overline{U}(z).
\end{equation}
From where we obtain the equation,
\begin{equation}\label{sch}
\left\{\frac{d^2}{dz^2}+\overline{V}_p(z)\right\}\overline{U}(z)=-m^2\overline{\psi},
\end{equation}
where
\begin{equation}
\overline{V}_p(z)=e^{2A_p}[\frac{1}{4}(A'_p)^2+\frac{1}{2}(A''_p)].
\end{equation}
We can not write the equation above in a supersymmetric form. Then
we can not exclude the existence of tachyonic modes. On the other
hand, we can find  a numerical solution for the function
$\overline{U}(y)$ in the equation (\ref{motion}). In this way we can
analyze the details of this solution and its contribution in the
effective action  for the massive modes of the tensor gauge field. The function ${\overline{U}(y)}^2
e^{-2A_p(y)}$, which is important in the effective action (27), it is plotted in Fig.(\ref{wavediv}). We can conclude that the effective action is not finite. Therefore,
there is no localized massive mode for the Kalb-Ramond field.

\section{Dilatonic deformed brane}

In the last section, we have not found signals of existence of zero
modes or massive modes trapped to the deformed membrane. The
coupling between the membrane (described by a two-bounce solution)
and the tensor gauge field is strictly due to the space-time metric.
Then, if we want to find localized modes we must modify the
structure of our membrane.

In this point, we would like following the procedure of Refs.\cite{kehagias,kazuo}, where the gauge field localization is
produced by including a new scalar field in the model: the dilaton.
By adding this field in the Einstein equations, we obtain a new
metric behavior and new information about the dynamics of the
membrane. The question here is to understand the behavior of the
Kalb-Ramond field in this new background.

The first step in this analysis is to study the Einstein equations
in this background. We get the action for the membrane, now with two
scalar fields \cite{kehagias},
\begin{equation}
S=\int d^{5}x
\sqrt{-G}[2M^{3}R-\frac{1}{2}(\partial\phi)^{2}-\frac{1}{2}%
(\partial\pi)^{2}-V_p(\phi,\pi)]
\end{equation}
where we denote by $\phi$ the scalar field responsible for the
membrane. The field $\pi$ represents the dilaton. It is assumed a
new ansatz for the spacetime metric:
\begin{equation}
ds^{2}=e^{2A(y)}\eta_{\mu\nu}dx^{\mu}dx^{\nu}+e^{2B(y)}dy^{2}.
\end{equation}
The equations of motion are given by
\begin{equation}
\frac{1}{2}(\phi^{\prime})^{2}+\frac{1}{2}(\pi^{\prime})^{2}-e^{2B(y)}V(%
\phi,\pi)=24M^{3}(A^{\prime})^{2},
\end{equation}
\begin{equation}
\frac{1}{2}(\phi^{\prime})^{2}+\frac{1}{2}(\pi^{\prime})^{2}+e^{2B(y)}V(%
\phi,\pi)=-12M^{3}A^{\prime\prime}-24M^{3}(A^{\prime})^{2}+12M^{3}A^{\prime}B^{%
\prime},
\end{equation}
\begin{equation}
\phi^{\prime\prime}+(4A^{\prime}-B^{\prime})\phi^{\prime}=\partial_{\phi}V,
\end{equation}
and
\begin{equation}
\pi^{\prime\prime}+(4A^{\prime}-B^{\prime})\pi^{\prime}=\partial_{\pi}V.
\end{equation}
To obtain the first order equations, we choose the following
superpotential $W_p(\phi)$ \cite{kehagias}:
\begin{equation}
V_p=e^{\frac{\pi}{\sqrt{12M^{3}}}}\{\frac{1}{2}\left(\frac{\partial
W_p}{\partial\phi}%
\right)^{2}-\frac{5M^{2}}{2}W_p(\phi)^{2}\}.
\end{equation}
The two kink solutions of the general form (\ref{twokink}) are used in Eq.(\ref{sup}) and we obtain:
\begin{equation}\label{sing}
\pi=-\sqrt{3M^{3}}A_p,
\end{equation}
\begin{equation}
B=\frac{A_p}{4}=-\frac{\pi}{4\sqrt{3M^{3}}},
\end{equation}
and
\begin{equation}
A_p^{\prime}=-\frac{W_p}{3}.
\end{equation}
Unfortunately, as we can see in Eq.(\ref{sing}), the solution for
the dilaton makes the space-time singular. The Ricci scalar for this
new geometry is now given by
\begin{equation}
R=-[8A_p''+18(A_p')^{2}]e^{\frac{\pi}{2\sqrt{3M^{3}}}}
\end{equation}

\begin{figure}{\centerline
{\epsfig{figure=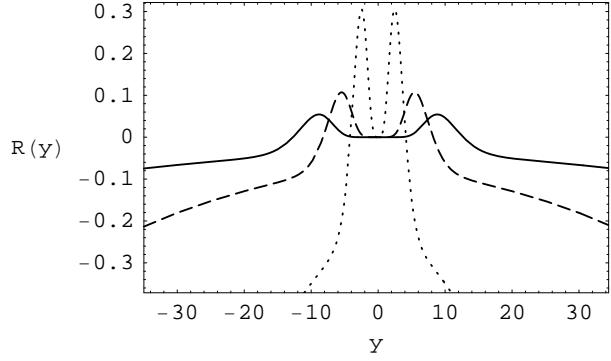,width=8cm,height=5cm} }
\caption{Plots of the solution of the curvature invariant
$R(y)$ with $p=3$ (doted line), $p=5$ (dashed line) and $p=7$ (solid
line).} \label{fig.3}}
\end{figure}

From the Fig.(\ref{fig.3}), there is a region near the membrane
where the Ricci scalar is null. When we take different values for
$p$, the width of this region increases due to the deformation
introduced. By another hand, because we have put the dilaton, the
curvature scalar decreases indefinitely in regions far from the
membrane.

\section{Kalb-Ramond field on dilatonic deformed brane}

Now we try again to localize the tensor gauge field, but in the new
background described in the section above. The issue here is to verify if
the dilaton coupling will be enough to localize the Kalb-Ramond
field in the deformed membrane. The dilaton coupling is introduced
in the model in the following way \cite{dilaton1,dilaton2}:
\begin{equation}
S\sim\int d^{5}x(e^{-\lambda\pi}H_{MNL}H^{MNL}).
\end{equation}
Therefore, we must analyze the equations of motion of the tensor
gauge field in the dilaton background. The new equation of motion
is:
\begin{equation}
\partial_{M}(\sqrt{-g}g^{MP}g^{NQ}g^{LR}e^{-\lambda\pi}H_{PQR})=0.
\end{equation}
With the gauge choice $B_{\alpha 5}=\partial\mu B^{\mu\nu}=0$ and
with the separation of variables
$B^{\mu\nu}(x^{\alpha},y)=b^{\mu\nu}(x^{\alpha})U(y)=b^{%
\mu\nu}(0)e^{ip_{\alpha}x^{\alpha}}U(y)$ where $p^{2}=-m^{2}$, it is
obtained a differential equation which give us information about the
extra dimension, namely
\begin{equation}\label{zero}
\frac{d^{2}U(y)}{dy^{2}}-(\lambda\pi^{\prime}(y)+B^{\prime}(y))\frac{dU(y)}{dy}=-m^{2}e^{2(B(y)-A(y))}U(y)
\end{equation}
For the zero mode, $m=0$, a particular solution of the equation
above is simply $U(y)\equiv cte$. This is enough for the following
discussion. The effective action for the zero mode in $D=5$ is
\begin{equation}
S\sim\int d^{5}x(e^{-\lambda\pi}H_{MNL}H^{MNL})=\int dy
U(y)^{2}e^{(-2A(y))+B(y)-\lambda\pi(y)}\int d^{4}x(h_{\mu\nu\alpha}h^{\mu\nu\alpha}).
\end{equation}
Given the solution $U(y)$ constant and regarding the solutions for
$A_p(y)$, $B(y)$ e $\pi(y)$, it is possible to show clearly that the
integral in the $y$ variable above is finite if $\lambda >
\frac{7}{4\sqrt{3M^3}}$, and  for $p$ finite. As a consequence, for
a specific value of the coupling constant $\lambda$ it is possible
to obtain a localized zero mode of the Kalb-Ramond field.

We should now consider a discussion about massive modes in this
background. For this, we must analyze the Eq. (\ref{zero}) for
$m\neq0$ trying to write it in a Schroedinger-like equation through
the following change
\begin{equation}\label{trans2}
y\rightarrow z=f(y),\,\,\,\,\,\,  U=\Omega\overline{U},
\end{equation}
with
\begin{equation}
\Omega=e^{\left(\frac{\alpha}{2}+\frac{3}{8}\right)A}, \,\,\,\,\,\, \frac{dz}{dy}=e^{-\frac{3}{4}A},
\end{equation}
were,
\begin{equation}
\alpha=\frac{1}{4}-\sqrt{3M^3}\lambda.
\end{equation}
After all the necessary calculations we arrive at the equation we
want to analyze, namely
\begin{equation}\label{schro}
\left\{-\frac{d^2}{dz^2}+\overline{V}(z)\right\}\overline{U}=m^2\overline{U},
\end{equation}
where the potential $\overline{V}_p(z)$  assumes the form,
\begin{equation}\label{pot_reson}
\overline{V}(z)=e^{\frac{3}{2}A}\left[\left(\frac{\alpha^2}{4}-\frac{9}{64}\right)(A')^2-\left(\frac{\alpha}{2}+\frac{3}{8}\right)A''\right].
\end{equation}
We can write the potential in function of
the derivatives respect to $z$,
\begin{equation}\label{vp}
\overline{V}_p(z)=\left[\beta^2(\dot{A}_p)^2-\beta\ddot{A}_p\right],
\end{equation}
where,
\begin{equation}
\beta=\frac{\alpha}{2}+\frac{3}{8}.
\end{equation}

\begin{figure}{\centerline
{\epsfig{figure=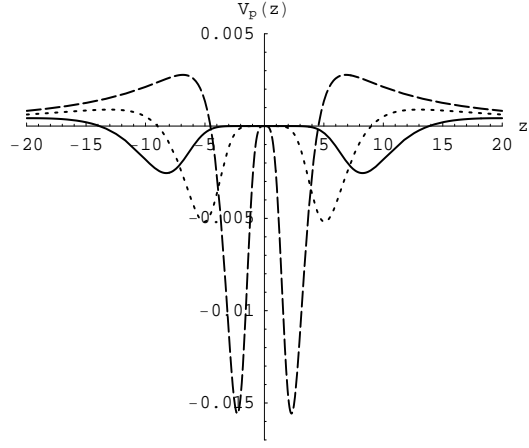,width=7cm,height=6cm} }
\caption{Plots of the potential $\overline{V}_p(z)$  with
$p=3$ (dashed line),
  $p=5$ (doted line) and $p=7$ (solid line). We have put $\sqrt{3M^3}\lambda=2$. } \label{fig.30}}
\end{figure}

We can see from the Fig.(\ref{fig.30}) that the potential is
affected by the deformation procedure introduced in this work. We
identify the existence of two minima whose distance increases when
we increase the values of $p$. The form of the potential is also directly
changed. Note that we use $\sqrt{3M^3}\lambda > 1$  in order to
obtain a potential like (\ref{vp}), i.e., the standard form found
when we write Schroedinger-like equations. This choice is
fundamental in order to find finite results regarding the behavior
of the Kalb-Ramond field.

It is interesting to point out that the Schrodinger-type equation
(\ref{schro}) can be written in the supersymmetric quantum mechanics
scenario as follows,
\begin{equation}\label{susy_qm}
Q^{\dag} \, Q \,
\overline{U}(z)=\left\{\frac{d}{dz}-\beta\dot{A}\right\}\left\{\frac{d}{dz}+\beta\dot{A}\right\}\overline{U}(z)=-m^2\overline{U}(z).
\end{equation}
From the form of the Eq. (\ref{susy_qm}), we exclude the possibility
of normalized negative energy modes to exist. On the other hand, we
exclude also the possibility of the presence of tachyonic modes,
which is a necessary condition to keep the stability of
gravitational background.

\begin{figure}{\centerline{
\epsfig{figure=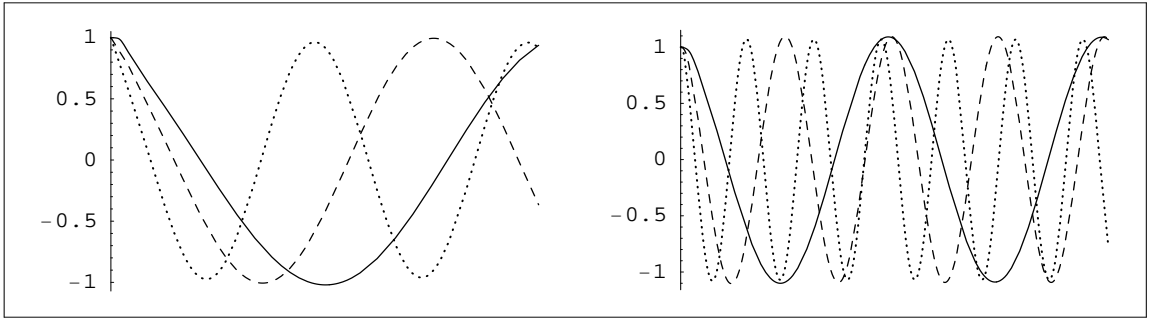,width=15.2cm,height=4.2cm}}
\caption{ Plots of $\overline{U}_p(z)$ for $p=1$ (doted
line), $p=3$ (dashed line), $p=5$ (solid line) and $p=7$ (thick
line), where we put $m^2 \leq {\overline{V}_p}(z)_{max}$ (left) and
$m^2 > {\overline{V}_p}(z)_{max}$ (right)}\label{wavepdil}}
\end{figure}

We cannot find analytical solution of the massive modes wave
function in Schrodinger equation. However we will be able to analyze
the solution for $\overline{U}_p$ by numerically solving the
equation (\ref{schro}). We plot in Fig.(\ref{wavepdil}) the wave
function so obtained for two values of $m^2$. As we can observe,
when we make $m^2 > \overline{V}_p(z)_{max}$, we minimize the
contribution due to the deformations over the solution
$\overline{U}_p$. However, regarding $m^2 \leq
{\overline{V}_p}(z)_{max}$, as we increase $p$ we reduce the
frequency of oscillations of the solutions $\overline{U}_p$. We must
remember that the search for finite solutions it was only possible due
to the choice $\sqrt{3M^3}\lambda> 1$.

As mentioned in Ref. \cite{adauto}, the behavior of the
wave function suggests us a free motion in the bulk, but no trapping
in the membrane.

\begin{figure}{\centerline{ \epsfig{figure=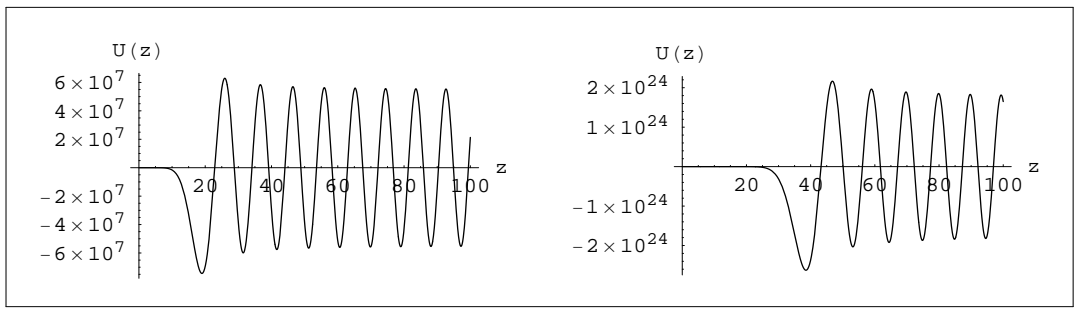,width=14.2cm,height=4cm}},\caption{Plots of $\overline{U}_p(z)$ for $p=1$, where we put $m=0,5$, $\sqrt{3M^3}\lambda=20$ (left) and $\sqrt{3M^3}\lambda=40$ (right)}. \label{dil}}
\end{figure}

An interesting point to be investigated is how the intensity of the
dilaton constant coupling may modify the patterns of solutions
observed in Fig.(\ref{wavepdil}). For this, we solve again the
equation (\ref{schro}), but this time, changing the values of the
constant coupling $\lambda$. We plot in Fig.(\ref{dil}) the
function $\overline{U}_p(z)$, this time for $\sqrt{3M^3}\lambda=20$
on the left and $\sqrt{3M^3}\lambda=40$ on the right. We note the
suppression of the mode oscillations in regions near the membrane
due to the increasing of $\lambda$. On the other hand, in regions far
away from $z=0$ the amplitude oscillations grows.

\begin{figure}{ \epsfig{figure=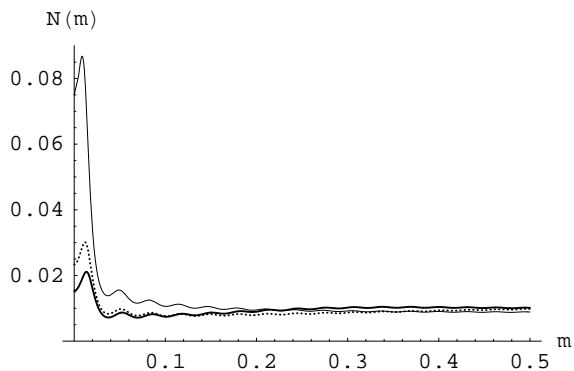,width=8cm}
\caption{Plots of $M_p(m)$ for $p=1$ (thin line), $p=3$ (points) and $p=5$ (thick line). }\label{resop}}
\end{figure}

\begin{figure}{\centerline{ \epsfig{figure=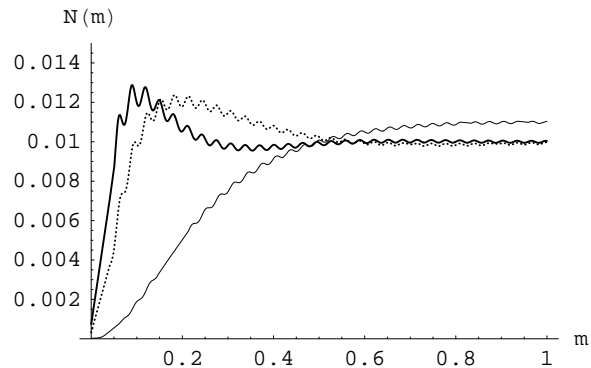,width=8cm}} \caption{Plots of $M_p(m)$ for $p=1$ (thin line), $p=3$ (points) and $p=5$ (thick line).}\label{reso2p}}
\end{figure}

In order to better understand the coupling between massive modes and
matter on the membrane we should know, starting from Eq.(\ref{susy_qm}), the amplitude of the plane wave function
$\overline{U}_p(z)$ normalized at $z=0$. The quantity
$|\zeta\overline{U}_p(0)|^2$, being $\zeta$ a normalization
constant, should give us the probability of finding a mode of mass
$m$ at $z=0$. In the Fig.(\ref{resop}) we plot
$M_p(m)=|\zeta\overline{U}_p(0)|^2$ where we can identify, for
$p=1$, a resonant peak near $m=0$, precisely for $m=9\times
10^{-3}$. We may interpret that, in this case, the probability of
finding light modes or massless modes coupled to the membrane is
bigger than for heavier modes. This characteristics disappears when
we change the values of $p$. As we can observe in Fig.(\ref{resop}), the resonant structure tends to disappear in
accordance to the results of localization of the zero mode.  We can
still test the consistency of the above results regarding again the
model without the dilaton coupling, sections 4.1 and 4.2. In this
case, we do not find signals of localization of the Kalb-Ramond
field. In this way, we can extract the function $M_p(m)$ from
equation (\ref{sch}) by the same steps discussed before and plot the
results in Fig.(\ref{reso2p}). As we expect, the resonant
structure disappears and the couplings of the zero modes is highly
suppressed.

\section{Conclusions}

In this article we analyze under several aspects the localization
properties of a real scalar field and the Kalb-Ramond tensor gauge
field in a very specific type of membrane: the deformed membrane.

We have obtained a scalar zero mode whose existence obeys to a
restriction due to the deformation structure included in the
analysis. When we consider massive modes, we obtain through a Schrodinger
-like equation an effective potential different of that found in the
usual models of membranes of the kink type. In the spectrum of
massive states we have found a resonance near the membrane through
the probability function. This resonance disappears when we deform
more and more the membrane, i.e., we increase the number $p$ in the
membrane solutions. In general, we can conclude that the zero mode coupling to the membrane is bigger than the massive mode coupling one.

The analysis of the Kalb-Ramond field it is jeopardized since the effective action is not normalizable: we have not zero modes for the Kalb-Ramond field. The resulting equation of motion for the massive modes is not found and it can not be written in a form of Schrodinger-like equation. This fact do not allows us to interpret quantum mechanically the problem. What we do to circumvent this result is to add one more field in the model, the dilaton, and this changes a little the gravitational background. After this modification, we can, under some conditions, find a localized tensorial zero mode. Related to the spectra of massive states, we see that the effective potential in the Schrodinger-like equation is affected by the deformations made in the membranes. The numerical analysis of that equation for massive states reveals that there are plane waves describing the free propagation of particles in the bulk. The dilaton coupling change the amplitude of oscillations of the modes away from the membrane. Indeed, studying the coupling of the matter massive states with the membrane we have found a resonance, which again disappears with the deformations. The resonance structures show us that only light modes of the KK spectrum present not suppressed coupling with the membrane. Finally, we showed the consistency of the results obtained with those from the model without the dilaton.

The authors would like to thank Funda\c{c}\~{a}o Cearense de apoio ao Desenvolvimento
Cient\'{\i}fico e Tecnol\'{o}gico (FUNCAP) and Conselho Nacional de Desenvolvimento
Cient\'{\i}fico e Tecnol\'{o}gico (CNPq) for financial support.



\end{document}